\documentclass[conference]{IEEEtran}
\IEEEoverridecommandlockouts
\usepackage{url}
\usepackage{multicol}
\usepackage{multirow}
\usepackage{amssymb}
\usepackage{pifont}
\usepackage{amsmath,amssymb,amsfonts}
\usepackage{amsmath,amssymb,amsfonts,color,soul}
\usepackage[ruled,vlined,linesnumbered]{algorithm2e}
\usepackage[flushleft]{threeparttable}
\usepackage{graphicx}
\usepackage{textcomp}
\usepackage{xcolor}

\def\BibTeX{{\rm B\kern-.05em{\sc i\kern-.025em b}\kern-.08em
    T\kern-.1667em\lower.7ex\hbox{E}\kern-.125emX}}

\usepackage[nolist]{acronym}
\acused{wifi}
\acused{wlan}
\acused{gnss}

\usepackage[backend=biber,style=ieee,  sortcites=true,isbn=false, doi=false, url=false,mincitenames=1, maxcitenames=1,maxbibnames=3,hyperref=false]{biblatex}
\usepackage{balance}

\usepackage{soul}
\soulregister\cite7
\soulregister\ref7
\soulregister\SI7
\soulregister\ac7

\usepackage[detect-all=true]{siunitx}
\DeclareSIUnit\decibelm{dBm}
    
\usepackage{array}
\usepackage{booktabs}
\usepackage{mathtools}

\DeclarePairedDelimiter\abs{\lvert}{\rvert}%

\newcommand{\MYheader}{20th International Conference on Pervasive Computing and Communications (PerCom 2022)
March 21--25, 2022, Pisa, Italy.}
\newcommand{\MYcopyrigth}{XXX-1-7281-6455-7/20/\$31.00~\copyright~2020~IEEE\hfill}

\renewcommand{\MYheader}{}
\renewcommand{\MYcopyrigth}{}

\makeatletter
\def\ps@headings{%
	\def\@oddhead{}
	\def\@evenhead{}
	\def\@oddfoot{}%
	\def\@evenfoot{}}
\def\ps@IEEEtitlepagestyle{%
	\def\@oddhead{\hfill\MYheader\hfill}
	\def\@evenhead{\hfill\MYheader\hfill}
	\def\@oddfoot{\MYcopyrigth}%
	\def\@evenfoot{\MYcopyrigth}
    }
\makeatother
\begin{document}

\title{Data Cleansing for Indoor Positioning Wi-Fi Fingerprinting Datasets

\thanks{%
Corresponding Author: D. Quezada Gaibor (\texttt{quezada@uji.com})
}
\thanks{The authors gratefully acknowledge funding from European Union’s Horizon 2020 Research and Innovation programme under the Marie Sk\l{}odowska Curie grant agreements No.~$813278$ (A-WEAR: A network for dynamic wearable applications with privacy constraints, {http://www.a-wear.eu/}) and No.~$101023072$ (ORIENTATE: Low-cost Reliable Indoor Positioning in Smart Factories, {http://orientate.dsi.uminho.pt}).}
}

\author{%
\IEEEauthorblockN{%
Darwin Quezada-Gaibor%
\IEEEauthorrefmark{1}\textsuperscript{,}\IEEEauthorrefmark{2}, %
Lucie Klus%
\IEEEauthorrefmark{2}\textsuperscript{,}\IEEEauthorrefmark{1}, %
Joaquín Torres-Sospedra%
\IEEEauthorrefmark{3}, %
\\Elena Simona Lohan%
\IEEEauthorrefmark{2}, %
Jari Nurmi%
\IEEEauthorrefmark{2}, %
Carlos Granell%
\IEEEauthorrefmark{1},
    and Joaquín Huerta%
\IEEEauthorrefmark{1}%
}

\IEEEauthorblockA{\IEEEauthorrefmark{1}\textit{Institute of New Imaging Technologies}, \textit{Universitat Jaume I}, Castellón, Spain}
\IEEEauthorblockA{\IEEEauthorrefmark{2}\textit{Electrical Engineering Unit}, \textit{Tampere University}, Tampere, Finland}
\IEEEauthorblockA{\IEEEauthorrefmark{3}\textit{ALGORITMI Research Centre}, \textit{Universidade do Minho}, Guimarães, Portugal}
}

\maketitle

\begin{abstract}
Wearable and IoT devices requiring positioning and localisation services grow in number exponentially every year. This rapid growth also produces millions of data entries that need to be pre-processed prior to being used in any indoor positioning system to ensure the data quality and provide a high Quality of Service (QoS) to the end-user. In this paper, we offer a novel and straightforward data cleansing algorithm for WLAN fingerprinting radio maps. This algorithm is based on the correlation among fingerprints using the Received Signal Strength (RSS) values and the Access Points (APs)'s identifier. We use those to compute the correlation among all samples in the dataset and remove fingerprints with low level of correlation from the dataset. We evaluated the proposed method on $14$ independent publicly-available datasets. As a result, an average of $14\%$ of fingerprints were removed from the datasets. The 2D positioning error was reduced by $2.7\%$ and 3D positioning error by $5.3\%$ with a slight increase in the floor hit rate by $1.2\%$ on average. Consequently, the average speed of position prediction was also increased by $14\%$.
\end{abstract}


\begin{acronym}[XXX] 
\acro{ap}[AP]{Access Point}
\acro{ble}[BLE]{Bluetooth Low Energy}
\acro{cdf}[CDF]{Cumulative Distribution Function}
\acro{csi}[CSI]{Channel State Information}
\acro{fp}[FP]{fingerprinting}
\acro{fpc}[FPC]{fingerprinting clustering}
\acro{iot}[IoT]{Internet of Things}
\acro{ips}[IPS]{Indoor Positioning System}
\acro{kde}[KDE]{Kernel Density Estimation}
\acro{knn}[$k$-NN]{$k$-Nearest Neighbors}
\acro{lda}[LDA]{Linear Discriminant Analysis}
\acro{lbs}[LBS]{location-based service}
\acro{los}[LoS]{Line-of-Sight}
\acro{mac}[MAC]{Media Access Control}
\acro{nn}[NN]{Nearest Neighbour}
\acro{pca}[PCA]{Principal Component Analysis}
\acro{qoe}[QoE]{Quality of Experience}
\acro{relu}[ReLU]{rectified linear}
\acro{rf}[RF]{Radio Frequency}
\acro{rp}[RP]{Reference Point}
\acro{rss}[RSS]{Received Signal Strength}
\acro{rssi}[RSSI]{Received Signal Strength Indicator}
\acro{nlos}[NLOS]{non-line-of-sight}
\acro{tsne}[t-SNE]{T-distributed Stochastic Neighbor Embedding}
\acro{ue}[UE]{User Equipment}
\acro{uwb}[UWB]{ultra-wideband}
\acro{vlc}[VLC]{Visible light communication}
\acro{wifi}[\mbox{Wi-Fi}]{IEEE 802.11 Wireless LAN}
\acro{wknn}[WKNN]{weighted k-nearest
neighbor}
\acro{wlan}[WLAN]{Wireless LAN}
\acro{wsn}[WSN]{Wireless Sensors Networks}

\end{acronym}

\section{Introduction}
\label{sec:introduction}

Indoor positioning and localization services are becoming increasingly demanded in various applications, including patient monitoring, ambient assisted living, smart parking assistance and indoor navigation apps. \ac{wifi}-based deployments are one of the most commonly used infrastructures for \ac{ips}~\cite{ometov2021survey}, mainly due to the global availability of \ac{wifi} \ac{ap}s and their standardized characteristics compliant with IEEE $802.11$, ensuring a good generalization properties across deployments. The measurements of \ac{wifi} \ac{rss} are easily obtainable by any \ac{ue}, ranging from mobile phones to battery-restricted \ac{iot} devices such as wearables. The main advantages of utilizing \ac{rss}-based ﬁngerprinting include its capability to perform well in environments with rich scattering characteristics and limited \ac{los} availability, in which the deterministic path-loss models usually fail~\cite{subedi2020survey}.

Fingerprinting is a simple technique, the position of a fingerprint (array of \ac{rss} measurements) can be estimated using the positions of the closest matches from a dataset with pre-recorded fingerprints (i.e., the radio map).
The radio map acquisition, pre-processing, training the matching algorithm and its optimization are referred to as the offline phase of fingerprinting. The online phase consists of finding the coordinates of the newly measured fingerprint in a real time.

The achievable positioning performance of the fingerprinting method depends on the scenario and strategy to collect the radio map. 
The localization algorithm, whether the \ac{knn}, or any alternative, can only fine-tune the positioning, which the training radio map allows it to.

In this work, we focus on improving the quality of the radio map by proposing a data cleansing scheme that is designed to remove the outlier samples from the radio map. The cleansing method calculates the similarity of each sample to the rest of the database based on the detected \ac{ap}s and their signal strength levels and removes the samples dissimilar to the rest. We then evaluate the proposed method on $14$ publicly available \ac{wifi} fingerprinting datasets and show they remain statistically unchanged. We also perform the fingerprinting-based positioning and show the improved performance of the cleansed databases when compared to the original ones.

The main contributions of this paper are as follows:
\begin{itemize}
    \item We propose a novel and straightforward algorithm for removing unnecessary samples from \ac{wifi} fingerprinting radio maps.
    \item We evaluate the proposed method and its capabilities on $14$ independent open-access datasets.
    \item We show, that the proposed method not only reduces the size of the datasets, but also improves the building hit, floor hit and positioning accuracy, on average, across all available datasets. Moreover, it reduces the time required to perform the user positioning. 
\end{itemize}

The rest of the paper is structured as follows. In Section~\ref{sec:related_work}, we discuss the related literature and works connected to our research. Section~\ref{sec:model} introduces the proposed data cleaning approach, which is later evaluated in Section~\ref{sec:usecases}. Additional impacting factors and things to consider are further elaborated in Section~\ref{sec:discussion} and the work is concluded in Section~\ref{sec:conclusions}.
\section{Related work}
\label{sec:related_work}

In this section, we discuss the related literature and outline other data cleaning methods focused on \ac{wifi}-based fingerprinting datasets. We also discuss the differences to our work and introduce its main advantages over the current State-of-the-Art.

Indoor positioning using \ac{wifi} \ac{rss} fingerprinting was broadly addressed across literature, most commonly considering \ac{knn}~\cite{torres2020comprehensive,pham2019improved} or various kinds of neural networks~\cite{klus2022transfer, abid2021improved} as the matching algorithm. Frequently, the individual works consider data pre-processing techniques, such as augmenting the radio map’s data representation~\cite{torres2020comprehensive}, reducing the number of \ac{ap}s by either removing the redundant ones~\cite{eisa2013removing}, applying radio map compression~\cite{klus2020rss, talvitie2017method}, or reducing the number of considered samples in the database by e.g., clustering~\cite{torres2020comprehensive, zhou2014indoor}.
Nevertheless, improving the quality of the database itself by performing data cleansing was hardly ever addressed. In this work, we evaluate the relevance of each sample in the training database (radio map) and remove the redundant ones. 

An example of localization dataset cleaning was proposed  in~\cite{lin2020locater}. There, the unlabelled fingerprint was first complemented with additional measurements, in the second iteration the coarse localization was realized, while in the last iteration the probabilistic model predicted the fine location. The work presented improved positioning results, but does not address the question of outliers within the positioning dataset.

The authors of~\cite{talvitie2014effect} studied the effect of coverage gaps in the \ac{rss} positioning datasets by artificially decreasing the database's positioning capabilities. The work showed that removing the samples from the dataset with uniform probability does not have strong diminishing effect. Alternatively, creating the measurement gaps in the training database strongly harmed the overall positioning performance in the deployment. Compared to this work, we eliminate the specific measurements from the database to boost the performance.

Simultaneous localization, outlier detection, and radio map interpolation was realized in~\cite{khalajmehrabadi2016structured}, which organizes the \ac{ap}s based on their similarity. The work supplements the missing measurements in the fingerprint by interpolating the measured \ac{rss} from the neighboring \ac{ap}s. The outlier detection algorithm discards the irrelevant measurements caused by, e.g., adversary attacks. The proposed Group-Sparsity localization system is able to perform even with the reduced database, but the only benchmark utilized in the comparison was compressive sensing, which is not commonly deployed in indoor localization schemes.

The authors of~\cite{9531633} identified several ways to enhance the radio map, including data cleansing and denoising. In~\cite{rs11111293}, the \ac{rssi} measurements were extracted to overcome sparsity with a stacked Denoising AutoEncoder (DAE). In~\cite{9205227}, denoising relied on another neural network architecture which handled not only sparsity but also \ac{rssi} fluctuations.  In~\cite{9462839}, denoising focused on learning the noise characteristics rather than the original characteristics. 

The challenge of missing and false values in crowdsensed \ac{rssi} sequence data was addressed in~\cite{sun2021data}. The mapping of \ac{rssi} sequences to the floor plan effectively boosts positioning capabilities, yet in many cases, as in this work, the temporal dependencies between samples are not available. Consequently, the conclusions from~\cite{sun2021data} cannot be applied directly.

The authors of \cite{stoyanova2010modeling} empirically determined the relation between the \ac{rss} data and its deviation. The study models the uncertainties in both static and mobile \ac{ue} situations, but restricts itself to the unobstructed link between the transmitter and the receiver. Nevertheless, uncertainty modelling and its estimation within the fingerprints can enhance the positioning model's knowledge and thus positively impact the positioning accuracy itself.

Compared to the works presented above, we restrict the dataset cleansing approach to directly remove the redundant, irrelevant and confusing samples from the training database, rather than finding the missing values and complementing the radio map, as is the case in many of the aforementioned references. By doing so, this work does not add any synthetically obtained information into the database, and therefore cannot introduce additional bias.

\section{Data cleansing}
\label{sec:model}
In this section, we provide a general overview of \ac{wifi} fingerprinting using the proposed data cleansing algorithm.

\subsection{Overview}
\ac{wlan} fingerprinting technique has been extensively researched during the last decade for both indoor and outdoor positioning, and it is being used in many commercial and open-source solutions. Generally, this technique consists of two phases - the online and the offline phase. In the offline phase \ac{rss} measurements are collected in known reference points to build a radio map. During the online phase, the \ac{rss} values collected in an unknown positions are matched with the fingerprints in the radio map using a matching algorithm such as \ac{knn} in order to estimate the device's position.

\vfill

\begin{figure}[ht]
    \centering
        \includegraphics[width=\columnwidth]{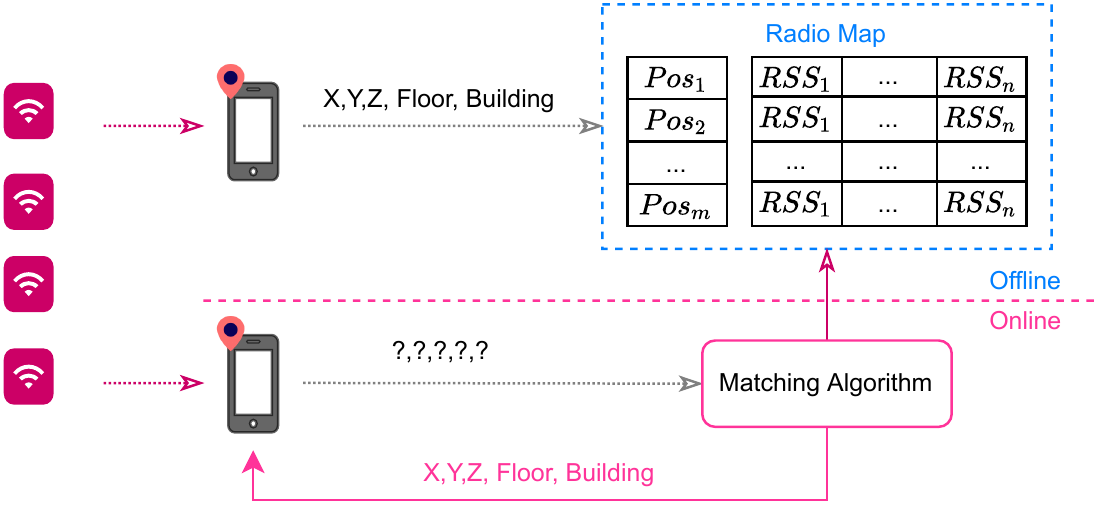}
    \caption{\ac{wifi} Fingerprinting Technique.}
    \label{fig:datasizelibs}
\end{figure}

\clearpage

Although this technique is considered one of the most robust techniques for indoor positioning, it may be affected by undesirable fluctuations in the signals, leading to an increase in terms of the positioning error. In case these fluctuations are not filtered out during the data collection in the offline phase, they might affect the positioning estimation in the online phase. Therefore, it is desirable to remove these noisy samples from the dataset in order to avoid errors in the position estimation and provide \ac{qoe} to the end-user.

In order to detect unnecessary fingerprints in the dataset, we propose a new straightforward method to remove outliers and/or unnecessary fingerprints in the radio map. This process consists of five steps detailed in the following paragraphs.

\subsection{Valid \ac{rss} values}
The first step is to determine the number of valid \ac{rss} values per fingerprint. In this case, the non-detected value ($\gamma$) has to be set in order to exclude it from the \ac{rss} values. Once the number of valid \ac{rss} values per fingerprint has been determined, the average or the maximum number of \ac{rss} values can be used to determine the correlation between samples. Considering a radio map $\Psi \in \mathcal{R}^{m \times n}$, where $m$ is the number of samples (fingerprints), and $n$ represents the number of \acp{ap} in the radio map, the average or the maximum number of valid \ac{rss} values ($\wp$) can be determined as follows:
\begin{align}
\label{eq:valrss}
\begin{split}
 \nu_{i} & = len(\Psi_i) | \forall \psi_{ij} \neq \gamma\\
 \wp & = \lfloor mean(\nu) \rfloor \text{ or } max(\nu)
\end{split}
\end{align}
where $\nu \in \mathcal{R}^{m}$ is a vector that contains the number of valid \ac{rss} values of the $i$-th sample, $i=1,2,3,...,m$ and $j=1,2,3,...,n$. $\psi_{ij}$ is the \ac{rss} value in the $i$-th and $j$-th position.

\subsection{Sort and Replace}
In this step, the \ac{rss} values are sorted in descending order and replaced for their corresponding \ac{ap} identifier. Then, the first $\wp$ columns are selected to compute the match percentage between the fingerprints.
\begin{align}
\label{eqn:eqlabel}
\begin{split}
 \mathcal{X}_{i} & = sort(\Psi_{i}, descending) \\
 x_{ij} & \leftarrow AP_j
\end{split}
\end{align}
where $\mathcal{X} \in \mathcal{R}^{m \times \wp}$ represents the radio map, using the \ac{ap} identifier instead of the \ac{rss} value and $x_{ij}$ is the \ac{rss} value in the $i$-th and $j$-th position. $AP_j$ is the \ac{ap} identifier in the $j$-th position.

\subsection{Compute the match percentage between samples}
The next step is to compute the correlation between samples. In this case, it is necessary to set a threshold ($\rho$) prior to computing the match percentage ($\Im$). This threshold ($\rho$) represents the minimal match percentage between samples.
The match percentage therefore is computed among all sample of the matrix $\mathcal{X}$.

$$
\mathcal{X}=
\begin{bmatrix}
x_{11}, x_{12}, & \hdots & x_{1\wp}\\
x_{21}, x_{22},  & \hdots & x_{2\wp}\\
\vdots & \ddots & \vdots\\
x_{m1}, x_{m2}, & \hdots & x_{m\wp}
\end{bmatrix}
$$

Thus, the $i$-th sample is compared with the $l$-th sample ($l=1,2,3,...,m$) under the following conditions: 

\begin{equation}
    \label{eq:conditions}
    \begin{aligned}
     \Im_i & =
     \begin{cases} 
        \Im_{i\_old}, & \text{if $\mathcal{X}_i=\mathcal{X}_l$}, \\
        \Im_{i\_old}, & \text{if $\Im_{i\_old} > \Im_{i}$}, \\
        \Im_{i\_old}, & \text{if $\Im_{i} < \rho$}, \\
        \frac{len(\mathcal{X}_i \cap \mathcal{X}_l )}{\wp}*100, & \text{otherwise}
    \end{cases}
    \end{aligned}
\end{equation}
where $\Im_{i\_old}$ is the previous match percentage computed between the $i$-th and the $l$-th sample ($\Im_{i}$).

\subsection{Remove unnecessary fingerprints}
In the last step of the proposed algorithm, all samples with zero match percentage are removed from the original radio.
These samples are considered outliers or unnecessary samples, given that they do not have high enough level of correlation with the rest of the samples. i.e., they may correspond to noisy samples poisoning the radio map.

\SetKwComment{Comment}{/* }{ */}
\SetKwInOut{KwIn}{Input}
\SetKwInOut{KwOut}{Output}
\begin{algorithm}
\caption{CleanDB}
\label{alg:cleandb}
\KwIn{X\_train, non\_detected\_value, threshold}
\KwOut{$\Psi_c$}
$\Psi \leftarrow X\_train$\\
$\Psi_c \leftarrow X\_train$\\
\tcc{Avg. or max. number of valid \ac{rss} values}
$\nu_{i} = len(\Psi_i) | \forall \psi_{ij} \neq \gamma$\\
$\wp = \lfloor mean(\nu) \rfloor \text{ or } max(\nu)$\\
\tcc{Sort and replace \ac{rss} values}
$\mathcal{X}_{i} = sort(\Psi_{i}, descending)$\\
$x_{ij} \leftarrow AP_j$\\
\tcc{Select the first $\wp$ columns of $\mathcal{X}$}
$\mathcal{X} \in \mathcal{R}^{m \times \wp}$\\
\tcc{Compute the match percentage}
 \For{i=1 to m}{
    \For{l=1 to m}{
    $\Im_i' = \frac{len(\mathcal{X}_i \cap \mathcal{X}_l )}{\wp}*100$\\
        \If{$\mathcal{X}_i \neq \mathcal{X}_l$ \&  $\Im_{i\_old} < \Im_{i}'$ \& $\Im_{i}' > \rho$}{
        $\Im_i = \Im_i'$
        }
    }
    $\Im_{i\_old} \leftarrow \Im_{i}$
}
\tcc{Remove samples with zero match percentage}
\For{i=1 to m}{
    \If{$\Im_{i} == 0$}{
    DEL ($\Psi_{ci}$)
    }
}
\end{algorithm}

Algorithm~\ref{alg:cleandb} summarizes all the previously explained steps to remove unnecessary fingerprints from the radio map. The algorithm requires three parameters, the training dataset, the non-detected value and the predefined match percentage ($\rho$). The output is the cleaned training dataset ($\Psi_c$).
\section{Experiments and Results}
\label{sec:usecases}

This section provides the experiment setup, a brief description of $14$ datasets used in the experiment, and the primary outputs of the proposed data cleansing algorithm for indoor positioning radio maps. The source code used in this experiment is available for public usage on Zenodo under the CC BY license~\cite{quezada_gaibor_darwin_2022_6381384}.

\subsection{Experiment setup}

The experiments were performed using a MacBook Pro with an M1 Pro chip packing a $10$-core CPU, a $16$-core GPU, and $16$GB of RAM. The software used for implementation was Python $3.9$, and these experiments were carried out using $14$ \ac{wifi} fingerprinting datasets collected in differing and heterogeneous scenarios. These datasets are:  UJI~1--2
, LIB~1--2 
(collected at Universitat Jaume I, Spain), MAN~1--2 
(collected at University of Mannheim, Germany), TUT~1--7 
(collected at Tampere University, Finland) and UTSIndoorLoc (collected at University of Technology Sydney, Australia)~~\cite{torres2020comprehensive,song2019cnnloc}. These datasets are representatives of multi-floor environments, all apart from MAN~1--2, which consist of measurements from one floor only. Additionally, UJI~1--2 datasets are apart from multi-floor environments also multi-building environments, as they consist of measurements obtained across several buildings. 

\begin{table*}[!hbtp]
    \tabcolsep 2.50pt

    \caption{Comparison $1$NN all data vs. $1$NN cleaned data}
    \label{table:results}
    \centering
    \begin{threeparttable}
    \begin{tabular}{
    l
    cccc
    ccccccccccc
    cccccc
    }
         \toprule
            &\multicolumn{4}{c}{Parameters}
            &\multicolumn{11}{c}{Baseline 1-NN}
            &\multicolumn{6}{c}{Cleaned DB + 1-NN}
   
            \\
         \cmidrule(rl){2-5}   
         \cmidrule(rl){6-16}
         \cmidrule(rl){17-22} 
            \multirow{2}{*}{Database}
            &\multirow{2}{*}{$\abs{\mathcal{T}_{TR}}$}
            &\multirow{2}{*}{$\abs{\mathcal{T}_{TE}}$}
            &\multirow{2}{*}{$\abs{\mathcal{A}}$}
            &\multirow{2}{*}{$\abs{\rho}$}
            
            &\multicolumn{1}{c}{$\zeta_{b}$}
            &\multicolumn{1}{c}{$\zeta_{f}$}
            &\multicolumn{1}{c}{{$\epsilon_{2D}$}}
            &\multicolumn{1}{c}{$\epsilon_{3D}$}
            &\multicolumn{1}{c}{$\delta$}
            &\multicolumn{1}{c}{$\tilde{\mathcal{T}}_{TR}$}
            &\multicolumn{1}{c}{$\tilde\zeta_{b}$}
            &\multicolumn{1}{c}{$\tilde\zeta_{f}$}
            &\multicolumn{1}{c}{$\tilde\epsilon_{2D}$}
            &\multicolumn{1}{c}{$\tilde\epsilon_{3D}$}
            &\multicolumn{1}{c}{$\tilde\delta$}

            &\multicolumn{1}{c}{$\tilde{\mathcal{T}}_{TR}$}
            &\multicolumn{1}{c}{$\tilde\zeta_{b}$}
            &\multicolumn{1}{c}{$\tilde\zeta_{f}$}
            &\multicolumn{1}{c}{$\tilde\epsilon_{2D}$}
            &\multicolumn{1}{c}{$\tilde\epsilon_{3D}$}
            &\multicolumn{1}{c}{$\tilde\delta$}
            \\
            
            &&&&
            &\multicolumn{1}{c}{$[\%]$}
            &\multicolumn{1}{c}{$[\%]$}
            &\multicolumn{1}{c}{$[\si{\meter}]$}
            &\multicolumn{1}{c}{$[\si{\meter}]$}
            &\multicolumn{1}{c}{$[\si{\second}]$}
            &\multicolumn{1}{c}{$[-]$}
            &\multicolumn{1}{c}{$[-]$}
            &\multicolumn{1}{c}{$[-]$}
            &\multicolumn{1}{c}{$[-]$}
            &\multicolumn{1}{c}{$[-]$}
            
            &\multicolumn{1}{c}{$[-]$}
            &\multicolumn{1}{c}{$[-]$}
            &\multicolumn{1}{c}{$[-]$}
            &\multicolumn{1}{c}{$[-]$}
            &\multicolumn{1}{c}{$[-]$}
            &\multicolumn{1}{c}{$[-]$}
            &\multicolumn{1}{c}{$[-]$}
            \\
         \midrule
LIB1 & 576 & 3120 & 174 & 33 & - & 99.84 & 3.035 & 3.043 & 0.531 & 1 & 1 & 1 & 1 & 1 & 1 & 0.844 & - & 1.000 & 0.998 & 0.998 & 0.843 \\ 

LIB2 & 576 & 3120 & 197 & 40 & - & 97.724 & 4.031 & 4.197 & 0.608 & 1 & 1 & 1 & 1 & 1 & 1 & 0.589 & - & 1.020 & 0.888 & 0.858 & 0.589 \\ 
MAN1 & 14300 & 460 & 28 & 34 & - & - & 2.877 & 2.877 & 0.376 & 1 & 1 & 1 & 1 & 1 & 1 & 0.961 & - & - & 0.981 & 0.981 & 0.914 \\ 

MAN2 & 1300 & 460 & 28 & 45 & - & - & 2.467 & 2.467 & 0.034 & 1 & 1 & 1 & 1 & 1 & 1 & 0.952 & - & - & 0.989 & 0.989 & 0.927 \\ 

TUT1 & 1476 & 490 & 309 & 35 & - & 90 & 8.623 & 9.601 & 0.401 & 1 & 1 & 1 & 1 & 1 & 1 & 0.674 & - & 1.014 & 0.903 & 0.873 & 0.769 \\ 

TUT2 & 584 & 176 & 354 & 30 & - & 72.727 & 11.218 & 12.893 & 0.073 & 1 & 1 & 1 & 1 & 1 & 1 & 0.664 & - & 1.039 & 0.964 & 0.939 & 0.660 \\ 

TUT3 & 697 & 3951 & 992 & 2 & - & 91.622 & 8.926 & 9.594 & 5.035 & 1 & 1 & 1 & 1 & 1 & 1 & 0.983 & - & 1.003 & 0.990 & 0.978 & 0.984 \\ 

TUT4 & 3951 & 697 & 992 & 1 & - & 95.265 & 6.152 & 6.406 & 5.424 & 1 & 1 & 1 & 1 & 1 & 1 & 0.996 & - & 1.000 & 0.998 & 0.999 & 0.992 \\ 

TUT5 & 446 & 982 & 489 & 21 & - & 88.391 & 6.387 & 6.924 & 0.393 & 1 & 1 & 1 & 1 & 1 & 1 & 0.798 & - & 1.001 & 0.969 & 0.956 & 0.809 \\ 

TUT6 & 3116 & 7269 & 652 & 2 & - & 99.986 & 1.959 & 1.959 & 27.612 & 1 & 1 & 1 & 1 & 1 & 1 & 0.997 & - & 1.000 & 0.990 & 0.990 & 0.997 \\ 

TUT7 & 2787 & 6504 & 801 & 0 & - & 99.185 & 2.110 & 2.351 & 27.429 & 1 & 1 & 1 & 1 & 1 & 1 & \ding{55} & - & \ding{55} & \ding{55} & \ding{55} & \ding{55} \\ 

UJI1 & 19861 & 1111 & 520 & 20 & 99.190 & 87.759 & 7.718 & 10.829 & 21.674 & 1 & 1 & 1 & 1 & 1 & 1 & 0.877 & 1.008 & 1.030 & 1.000 & 0.828 & 0.877 \\ 

UJI2 & 20972 & 5179 & 520 & 20 & 100.000 & 85.345 & 7.742 & 8.052 & 108.441 & 1 & 1 & 1 & 1 & 1 & 1 & 0.873 & 1.000 & 1.022 & 0.978 & 0.960 & 0.876 \\ 

UTS1 & 9108 & 388 & 589 & 20 & - & 92.784 & 7.769 & 8.757 & 4.076 & 1 & 1 & 1 & 1 & 1 & 1 & 0.923 & - & 1.008 & 1.002 & 0.962 & 0.888 \\ 
\midrule

Avg. & ~ & ~ & ~ & ~ & ~ & ~ & ~ & ~ & ~ & 1 & 1 & 1 & 1 & 1 & 1 & 0.856 & 1.004 & 1.012 & 0.973 & 0.947 & 0.856 \\ 

         \bottomrule
    \end{tabular}
    \end{threeparttable}
    \begin{tablenotes}
    \item ``-" indicates single building and/or floor. ``\ding{55}" represents the dataset where the cleansing algorithm was not able to find any unnecessary fingerprint.
    \end{tablenotes}
\end{table*}

The core algorithm to estimate the user or device position as well as to classify the fingerprints into buildings and floors was \acf{knn}. It was selected for its good positioning capabilities as previously demonstrated in the literature \cite{832252,ma2008cluster,torres2020comprehensive}. The hyperparameters set in the \ac{knn} algorithm are $k$ equal to $1$ and Manhattan distance as the distance metric to compute the similarity between the fingerprint vectors. The modules used are \textit{KNeighborsClassifier} and \textit{KNeighborsRegressor} from the \textit{Scikit-learn} (Sklearn) library. Additionally, positive data representation
~\cite{TORRESSOSPEDRA20159263} was used in all datasets prior to applying the proposed algorithm.

In order to choose the optimal threshold of the match percentage, the experiments were run using thresholds in intervals of $5\%$. If the positioning error increases or the floor hit rate decreases within the used interval, intermediate values are selected to run the algorithm. For this reason, there are thresholds of the match percentage equal to $1\%$, $2\%$, $20\%$, $21\%$, etc. (see Table~\ref{table:results}). The non-detected value used for all original datasets is \SI{100}{\decibelm} and the maximum number of valid \ac{rss} samples ($\wp$) is given by  Eq.~\ref{eq:valrss}. 

The results obtained with \ac{knn} using the original dataset and the cleansed dataset were compared in terms of mean 2D positioning error ($\epsilon_{2D}$), mean 3D positioning error ($\epsilon_{3D}$), building hit rate ($\zeta_{b}$), floor hit rate ($\zeta_{f}$), testing time ($\delta$) and the size of the training dataset ($\mathcal{T}_{TR}$). Given the heterogeneity of the datasets, the results obtained with the original dataset and the cleansed dataset were normalized in order to be compared, e.g., normalized mean 3D positioning error ($\tilde{\epsilon}_{3D}$). The values reported with the plain $1$-NN for the above mentioned metrics have been selected for the normalisation. i.e., normalised values will be relative to that baseline. 

\subsection{Results}

Table~\ref{table:results} shows the parameters of each dataset and the main results after running the \ac{knn} algorithm with the original datasets and with the cleansed datasets. $\abs{\mathcal{T}_{TR}}$ represents the number of training samples in the dataset, $\abs{\mathcal{T}_{TE}}$ is the number of testing samples and $\abs{\mathcal{A}}$ is the number of \acp{ap} in the dataset. 

For the baseline method, the 1-NN algorithm, the absolute and normalised values are provided.
After using the proposed data cleansing algorithm, most of the training datasets reduced their number of samples, with the exception of TUT~7 dataset in which the cleansing algorithm was not able to detect unnecessary samples. That is why the threshold was set to $0\%$ for TUT~7 dataset. The minimal number of unnecessary fingerprints removed from the datasets were $9$ of $3117$ in TUT~6 dataset ($\approx 0.29\%$), and the maximum number of removed fingerprints was $237$ from the LIB~2 dataset ($\approx 41\%$ of the original dataset size). In any case, the positioning error, floor and building hit rate were not negatively affected.

\begin{figure}[ht]
    \centering
        \includegraphics[width=\columnwidth]{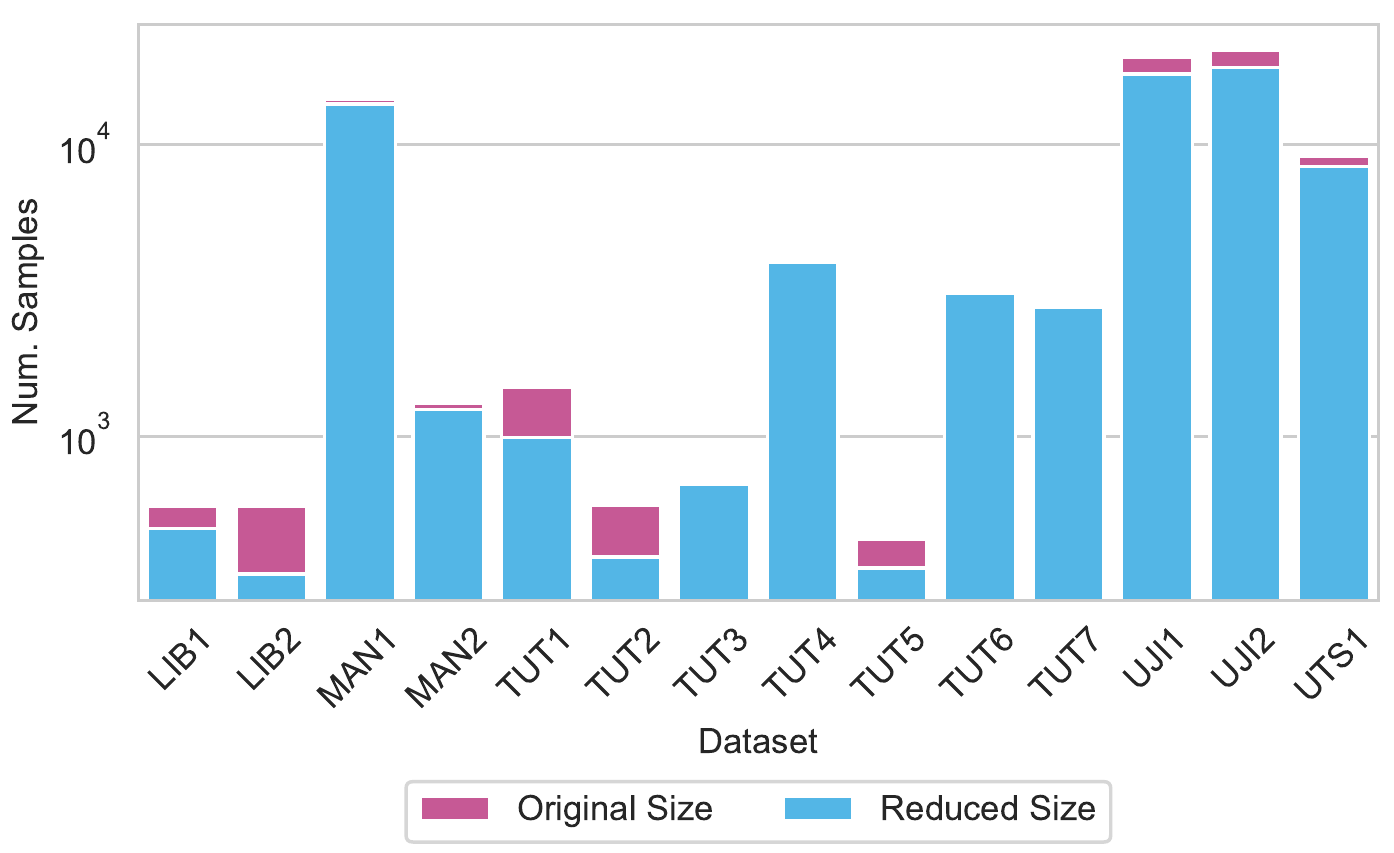}
    \caption{Dataset size before and after applying the data cleansing algorithm.}
    \label{fig:datasizelibs}
\end{figure}

Fig.~\ref{fig:datasizelibs} shows the number of fingerprints after (blue) and before the data cleansing (red). In all but one dataset the algorithm achieved at least a small reduction of their original size. Similarly, the prediction time was slightly reduced after applying the data cleansing algorithm by $\approx 14\%$ (see Table~\ref{table:results}).

Additionally, the use of the proposed data cleansing algorithm reveals a slight increment in the average building hit rate ($\tilde{\zeta}_b$) from $1$ to $1.004$ ($0.4\%$) and the average floor hit rate ($\tilde{\zeta}_f$) from $1$ to $1.012$ ($1.2\%$). Similarly, the proposed algorithm allowed us to reduce the positioning error in most of the datasets in both 2D ($\tilde{\epsilon}_{2D}$) and 3D ($\tilde{\epsilon}_{3D}$) positioning error. For instance, the normalized mean 3D positioning error in LIB~1 was reduced from $1$ to $0.998$ without affecting the floor hit rate. In LIB~2, the error was reduced from $1$ to $0.858$ ($\approx 58$cm), increasing the floor hit rate from $1$ to $1.020$ ($\approx 2\%$).

In general, the average normalized 2D positioning error decreased from $1$ to $0.973$ ($2.7\%$) and the average normalized 3D positioning error from $1$ to $0.947$ ($5.3\%$). The accuracy of the floor hit increased by $1.2\%$, and the building hit rate remained almost unchanged.

\begin{figure}[ht]
    \centering
        \includegraphics[width=\columnwidth]{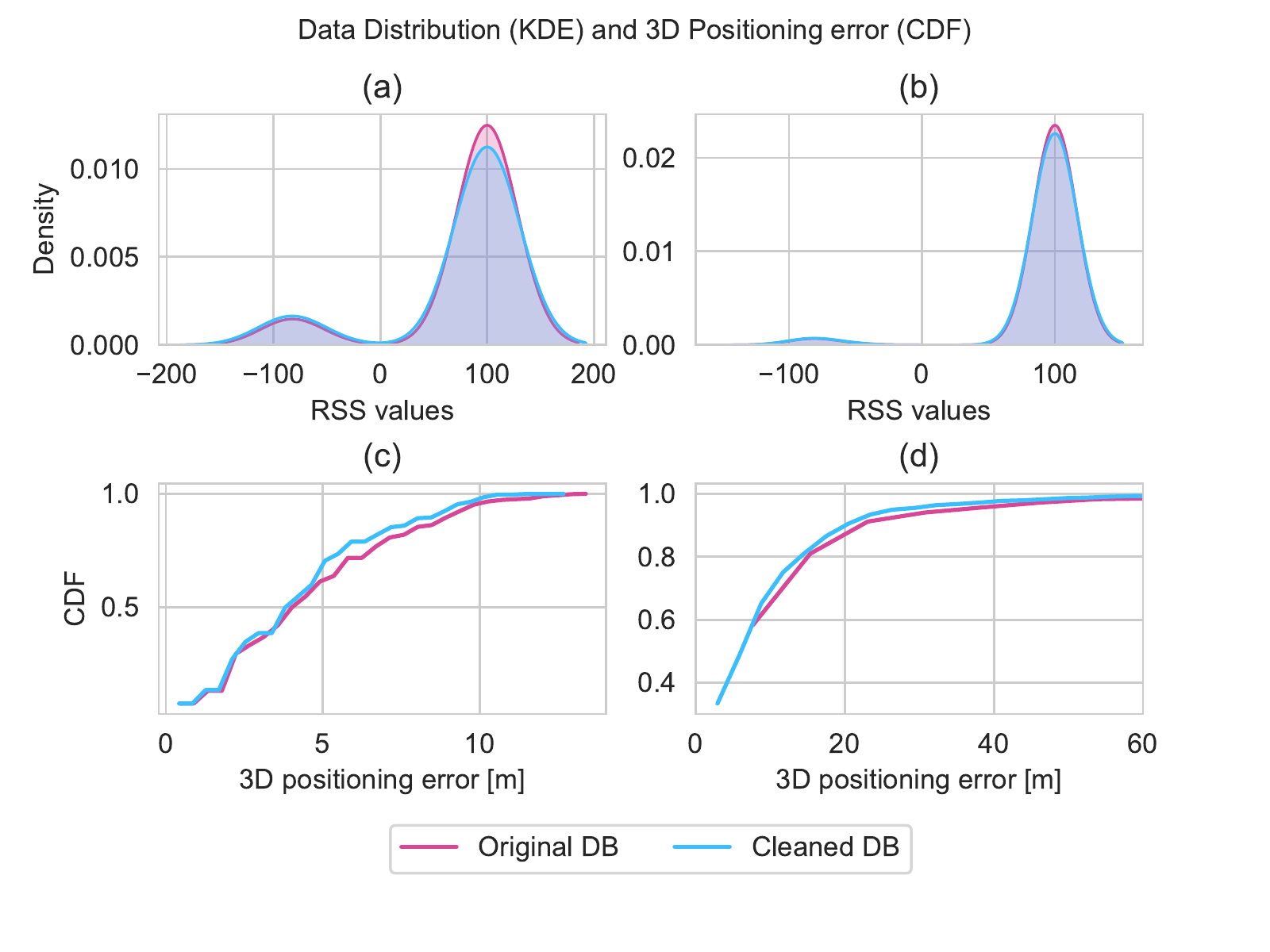}
    \caption{LIB~2 (a, c) and UJI~1 (b, d). Top (a--b): \ac{rss} distribution after and before the data cleansing. Bottom (c--d): CDF of the 3D positioning error.}
    \label{fig:densitycdf}
\end{figure}

Fig.~\ref{fig:densitycdf}a shows the distribution of the \ac{rss} values using \ac{kde} in LIB~2 dataset. \ac{kde} is used to visualize the data distribution and its density. The red colour denotes the distribution of the original dataset, whereas the blue colour represents the data distribution after applying the data cleansing. In this case, the vast majority of the values are non-detected values denoted by $100$. When the data cleansing algorithm is applied to the dataset, the data distribution is almost the same, but the density in the region of non-detected values was reduced, and the area of valid \ac{rss} values slightly increased. As can be observed from the \ac{cdf} plot (see Fig.~\ref{fig:densitycdf}c), the proposed data cleansing reduced the positioning error. 
For instance, the possibility of having a positioning error of less than $4 \si\metre$ is $54\%$ before the data cleaning and $60\%$ after it.

Similarly, Fig.~\ref{fig:densitycdf}b shows the distribution of the \ac{rss} values in UJI~1 dataset and the \ac{cdf} of the 3D positioning error in the same dataset (see Fig.~\ref{fig:densitycdf}d). UJI~1--2 are the only datasets with multiple buildings (3 buildings) and floors (4--5 floors). In UJI~1 dataset, we can observe errors over $100 \si\metre$, whereas the maximum positioning error obtained after the data cleansing is around $88 \si\metre$. The same pattern can also be observed in Table~\ref{table:results}, where the normalised mean 3D positioning error was reduced by $\approx 17\%$ ($\approx 2 \si\metre$).

\section{Discussion}
\label{sec:discussion}

Ensuring the quality of the data has become an essential step to provide better analysis and, therefore, better results. Indoor positioning datasets are not an exception; data collected from differing environments may contain irrelevant observations, outliers, missing data or noisy sample that may poison the radio map. Therefore, it is crucial to ``cleanse" the datasets to offer high-quality data to any model used to estimate the device position.

The proposed data cleansing algorithm offers a straightforward way of removing irrelevant fingerprints from indoor positioning radio maps without increasing the positioning error. In some cases, the proposed method also helps to provide a better position estimation, showing its potential for data cleansing in \ac{wifi} fingerprinting radio maps. However, the complexity of radio maps makes it difficult to detect irrelevant data or outliers in some datasets. For instance, in TUT~4 and TUT~6, the number of unnecessary fingerprints detected was insignificant compared to the size of the dataset. 

In the particular case of TUT~7, the proposed algorithm could not detect any unnecessary fingerprints. Even when the threshold was set with a minimal match percentage (less than $5\%$) between fingerprints, the positioning error was negatively affected. 

Although the average or the maximum number of valid \ac{rss} samples can be used in the proposed algorithm, the maximum number of valid \ac{rss} samples provides better performance than the average in some of the datasets. In some cases, both the average and the maximum can offer the same results but using different thresholds. For instance, in TUT~6 with $\rho$ equal to $5\%$ and average method can obtain the same positioning error as the one reported in Table~\ref{table:results}.

It is important to highlight that the proposed cleansing algorithm can be complemented with other tools or algorithms to remove unnecessary fingerprints from the radio map. 

\section{Conclusions}
\label{sec:conclusions}

In this paper, we offer a novel and straightforward algorithm to remove unnecessary samples from \ac{wifi} fingerprinting radio maps. This algorithm compares the \acp{ap} in common between fingerprints to compute the match percentage between each one under predefined conditions. The evaluation comprises $14$ multi-storey \ac{wifi} datasets taken with different strategies in different locations aiming at obtaining generalizable results. 

As a result, the proposed cleansing algorithm was able to remove unnecessary samples, reducing the size of the datasets by more than $14\%$, with an average improvement in the 2D positioning error of $2.7\%$ and $5.3\%$ in the 3D positioning error. Also, there was a slight improvement in the floor hit rate ($\approx~1.2\%$ on average). Additionally, the time required for position prediction was decreased by $14\%$. i.e., the proposed method is able to improve all metrics. 

Future work will analyze new techniques and algorithms to improve the quality of \ac{wlan} radio maps, combined with the proposed data cleansing algorithm.

\balance
\renewcommand*{\UrlFont}{\rmfamily}\printbibliography

\end{document}